\documentstyle[epsfig,aps,prb,multicol]{revtex}
\begin{document}

\title{Polarization forces in water deduced from single molecule data}

\author{E.V. Tsiper}

\address{
 School of Computational Sciences, George Mason University, Fairfax,
VA 22030\\
 Center for Computational Material Science, Naval Research Laboratory,
Washington, DC 20375
 \\etsiper@gmu.edu
}

\date{July 8, 2004}

\maketitle

\begin{multicols}{2}
 [ \begin{abstract}
 Intermolecular polarization interactions in water are determined
using a minimal atomic multipole model constructed with distributed
polarizabilities.  Hydrogen bonding and other properties of
water-water interactions are reproduced to fine detail by only three
multipoles $\mu_H$, $\mu_O$, and $\theta_O$ and two polarizabilities
$\alpha_O$ and $\alpha_H$, which characterize a single water molecule
and are deduced from single molecule data.
 \end{abstract} ]

Understanding polarization forces is crucial in many molecular systems
such as molecular clusters, liquids, or solids, specifically those
containing polar and polarizable molecules.  Polarization effects in
water are particularly strong, as can be judged by the enhancement of
the molecular dipole from 1.855 D for an isolated molecule to 2.6 ---
3.2 D in condensed state.\cite{iceDipole,parrinello} Water is a very
basic substance.\cite{gen} It is a fascinating object to study because
of its singular properties, its significance in biological systems,
and because it is a classic example of hydrogen bonding.\cite{Hbook}
Hydrogen bonding, which itself is one of the key elements of the
functioning of life, is largely a polarization effect.  Unfortunately,
no commonly accepted model describes it simply and accurately at the
same time.  Here we show that application of recent rules for minimal
atomic multipoles\cite{mame} combined with the notion of distributed
polarizabilities lead straightforwardly, without further intervention,
to a very transparent model for polarization forces in water.
Hydrogen bonding and other properties of water-water interactions are
reproduced to fine detail with only three atomic multipoles and two
polarizabilities, whose values are deduced based on single molecule
data.

Intermolecular potential for water has been extensively studied, with
about 150 models introduced since 1930s, indicating difficulties in
this area.\cite{kutepov} Recent accurate parameterizations involving
several tens of parameters are available based on tuning to rich
vibration-rotation-tunneling (VRT) spectra,\cite{VRT,VRTIII} or to
high-level quantum-chemical calculations,\cite{SAPT-5s} or
both.\cite{SAPT-5st} Some models are based on molecular multipole
moments and require high-order multipoles.\cite{batista} Following
seminal work by Rahman and Stillinger,\cite{stillinger} many empirical
models involve distributed
charges.\cite{finney,SPC,TIP5P,POL1,krogh,rick_A,POL5,lamoureux_A,ren_A,yu_A}
Most of the force fields use static charges thus ignoring or averaging
the polarization effects, while other models incorporate
polarizabilities
explicitly.\cite{POL1,krogh,rick_A,POL5,lamoureux_A,ren_A,yu_A} Work
\onlinecite{krogh} first introduced molecular polarizability of water
distributed over atomic sites.

It has been recently recognized that hydrogens need not be assigned
charges in distributed charge models.\cite{mame} The hydrogen's sole
electron participates in the chemical bond and is not centered at the
proton.  Therefore, hydrogen is best described by an atomic dipole
$\mu_H$ placed at the proton and directed along the bond.  Assigning
both charge and dipole causes redundancy and leads to unphysical
results.  This rule is an integral part of the minimal atomic
multipole expansion (MAME),\cite{mame} which eliminates the
redundancies by a careful choice of the minimal set of atomic
multipoles based on the Lewis structure of the molecule.

MAME rules lead to the following expression for the electrostatic
potential of a single water molecule:

\begin{eqnarray}
\phi({\bf r})&=&
 \mu_H\frac{({\bf r}-{\bf r}_1)\cdot{\bf r}_1/l}{|{\bf r}-{\bf r}_1|^3}
+\mu_H\frac{({\bf r}-{\bf r}_2)\cdot{\bf r}_2/l}{|{\bf r}-{\bf r}_2|^3}
  \nonumber\\
&+&\mu_O\frac{{\bf r}\cdot{\bf n}}
   {r^3}+\theta_O
\frac{2r^2-3({\bf r}\cdot{\bf n}_1)^2-3({\bf r}\cdot{\bf n}_2)^2}
  {2r^5}.
\label{phi}
\end{eqnarray}
 Since protons have no charge, neutrality allows no charge on the
oxygen either.  The dipole $\mu_O$ and quadrupole $\theta_O$ describe
the two lone pairs on oxygen.\cite{mame} Origin is at the oxygen,
${\bf r}_{1,2}$ are the positions of protons, $r_{1,2}=l$, ${\bf
n}=({\bf r}_1+{\bf r}_2)/|{\bf r}_1+{\bf r}_2|$ is the unit vector
along the symmetry axis, and ${\bf n}_{1,2}$ are unit vectors in the
directions of lone pairs (Fig.~1).  Experimental geometry has
$l=0.9572$ \AA\ and a nearly tetrahedral bond angle
$\beta=104.52^\circ$ between ${\bf r}_1$ and ${\bf
r}_2$.\cite{ExpGeom} We take ${\bf n}_1$ and ${\bf n}_2$ to be at the
tetrahedral angle $\beta^{'}=109.47^\circ$.\cite{POL5} Significant
deviation from this value leads to a dramatic deterioration of
accuracy of Eq.~(\ref{phi}) as seen in the inset.

 \centerline{\epsfig{file=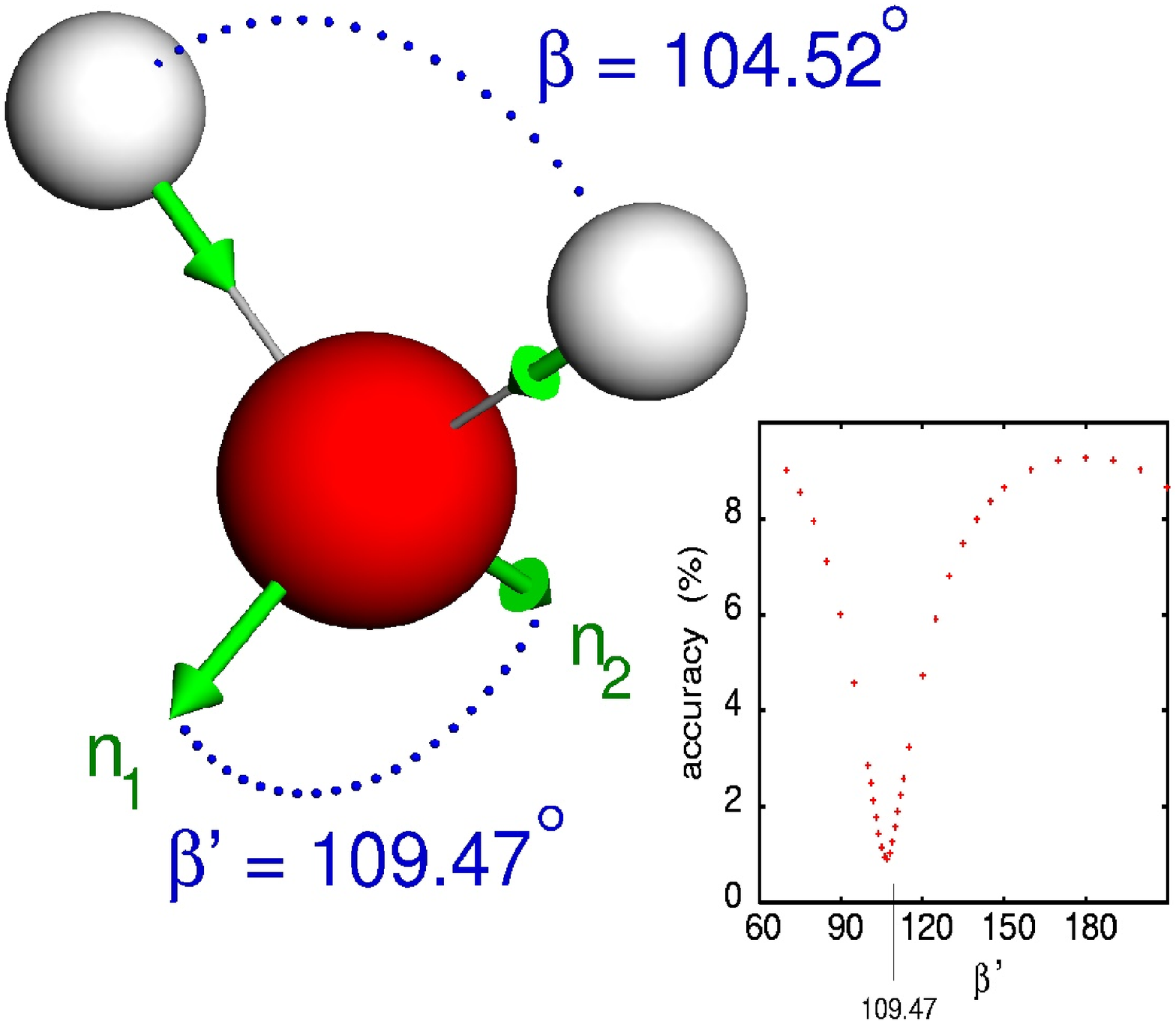,width=6.2cm}}
 \vskip 0.01 in
 \vbox{\baselineskip0.4 pt
 {\small {\bf Fig.~1} Geometry of a single water molecule.  Vectors
${\bf n}_1$ and ${\bf n}_2$ point in the direction of the lone pairs
on oxygen.  The inset shows an effect of varying $\beta^{'}$ on the
accuracy\cite{mame} of (\ref{phi}).  $\mu_H$, $\mu_O$ and $\theta_O$
are re-optimised for every $\beta^{'}$.  The vertical bar marks the
perfect tetrahedral angle. }}

Our goal is to extend the static model to describe the field induced
by a polarized molecule subject to external fields.  In doing so we
again keep only the minimal set of multipoles to avoid redundancies.
Charge redistribution part of the molecular polarizability\cite{pplus}
vanishes for water due to the absence of charged sites.  Thus we
assign polarizabilities to individual nuclei in such a way as to
reproduce experimental molecular polarizability.  The smallest
component is $\alpha_{yy}=1.4146(3)$ \AA$^3$ normal to the molecular
plane, the next is $\alpha_{zz}=1.4679(13)$ \AA$^3$ along the dipole
moment, and the largest is $\alpha_{xx}=1.5284(13)$ \AA$^3$ in the
longest dimension.\cite{ExpAlpha}

Atomic polarizabilities reflect the local atomic environments and need
not necessarily be isotropic.  Tetrahedral coordination of oxygen
suggests to assign it an isotropic polarizability $\alpha_O$.  For
hydrogens the polarizability $\alpha_H$ along the OH bond may differ
from the polarizability $\alpha_\bot$ normal to it.  To deduce
$\alpha_O$, $\alpha_H$, and $\alpha_\bot$ we express the molecular
polarizability,

\begin{eqnarray}
\alpha_{xx}&=&\alpha_O+2\alpha_H\sin^2\beta/2+2\alpha_\bot\cos^2\beta/2\nonumber,\\
\alpha_{yy}&=&\alpha_O+2\alpha_\bot,\nonumber\\
\alpha_{zz}&=&\alpha_O+2\alpha_H\cos^2\beta/2+2\alpha_\bot\sin^2\beta/2.
\label{alpha3}
\end{eqnarray}
 In a surprise twist, the determinant of this linear system is
identically zero.  Equations (\ref{alpha3}) are therefore dependent
and possess a solution only if the quantity

\begin{equation}
 \alpha_{xx}\cos^2\beta/2
+\alpha_{yy}(2\sin^2\beta/2-1)
-\alpha_{zz}\sin^2\beta/2
\label{identity}
\end{equation}
 is zero.  Thus, the model is adequate if the relation holds between
the molecular polarizability components.  Substituting the
experimental values into (\ref{identity}) we get 0.0093\AA$^3$, which
is indeed close to zero.  Two independent equations suggest that one
of the atomic polarizabilities can be safely omitted.  The natural
choice is to set $\alpha_\bot=0$, implying that the dipole moments on
protons can change their value, but not direction.  Solving
(\ref{alpha3}) we get

\begin{eqnarray}
\alpha_O&=&\alpha_{yy}=1.4146\ \text{\AA}^3,\ \ \ \ \text{and}\nonumber\\
\alpha_H&=&(\alpha_{xx}+\alpha_{zz})/2-\alpha_{yy}=0.0836\ \text{\AA}^3.
\label{alpha2}
\end{eqnarray}
 Thus, the bulk of molecular polarizability comes from the oxygen,
which is consistent with its atomic size, while the small
polarizabilities on the protons account for the (small) anisotropy of
the molecular polarizability tensor.

Three gas-phase multipoles from a density functional calculation,
$\mu_H=0.675$ D, $\mu_O=1.033$ D, and $\Theta_O=1.260$ D\AA\cite{mame}
result in the molecular dipole $\mu=1.854$ D and the quadrupole
components $\Theta=\Theta_{xx}-\Theta_{yy}=4.973$ D\AA,
$\Theta_{zz}=0.142$ D\AA.\cite{comment} These should be compared to
experimental data,\cite{Expmu,ExpQ} $\mu=1.8546(6)$ D,
$\Theta=5.126(25)$ D\AA, and $\Theta_{zz}=0.113(27)$ D\AA.

We again adjust the three atomic multipoles to satisfy the three
experimental values precisely to avoid any computational input.  The
molecular dipole and quadrupole are expressed in terms of the atomic
multipoles as

\begin{eqnarray}
\mu&=&\mu_O+2\mu_H\cos\beta/2,\nonumber\\
\Theta&=&6l\mu_H\sin^2\beta/2+3\theta_O\sin^2\beta^{'}/2,\nonumber\\
\Theta_{zz}&=&2l\mu_H(3\cos^2\beta/2-1)-\theta_O(3\cos^2\beta^{'}/2-1)
\label{static3}
\end{eqnarray}
 In practice, we face here an almost identical problem, in that the
determinant of (\ref{static3}) is small.  It becomes zero when an
ideal tetrahedral angle is substituted for $\beta$.  A relation
similar to $(\ref{identity})$ in this case reads simply
$\Theta_{zz}=0$.  Actual $\Theta_{zz}$ is indeed small, but not zero,
and $\beta$ deviates noticeably from $109.47^\circ$.  Nevertheless,
smallness of the determinant indicates that the finite accuracy data
can be satisfied by a range of atomic multipoles, and so the third
equation in (\ref{static3}) cannot be used reliably.

Thus, we use the first two equations to express $\mu_O$ and $\theta_O$
in terms of $\mu_H$, which guarantees to reproduce experimental $\mu$
and $\Theta$, while keeping reasonable $\Theta_{zz}$.  The DFT value
$\mu_H=0.675$ D yields $\mu_O=1.029$ D and $\theta_O=1.352$ D\AA, with
$\Theta_{zz}=0.160$ D\AA.  The model is thus completely defined and
readily yields the polarization energy $E_P$ for the water dimer,
trimer and larger clusters.\cite{pplus}

Water clusters from dimers on up have been extensively studied with
both experiment\cite{VRT,ExpGeomD,saykally_trimer,curtiss_1979} and
theory.\cite{VRT,VRTIII,MCY,stone_waterD,scoles_2001} Six-dimensional
adiabatic energy surface of the dimer has 8 equivalent
minima\cite{pople} split in a complex fashion by zero-point tunneling
motion.  Softness of the pair potential requires care when relating it
to the experimental observables.\cite{VRT}

 \vskip 0.01 in
 \centerline{\epsfig{file=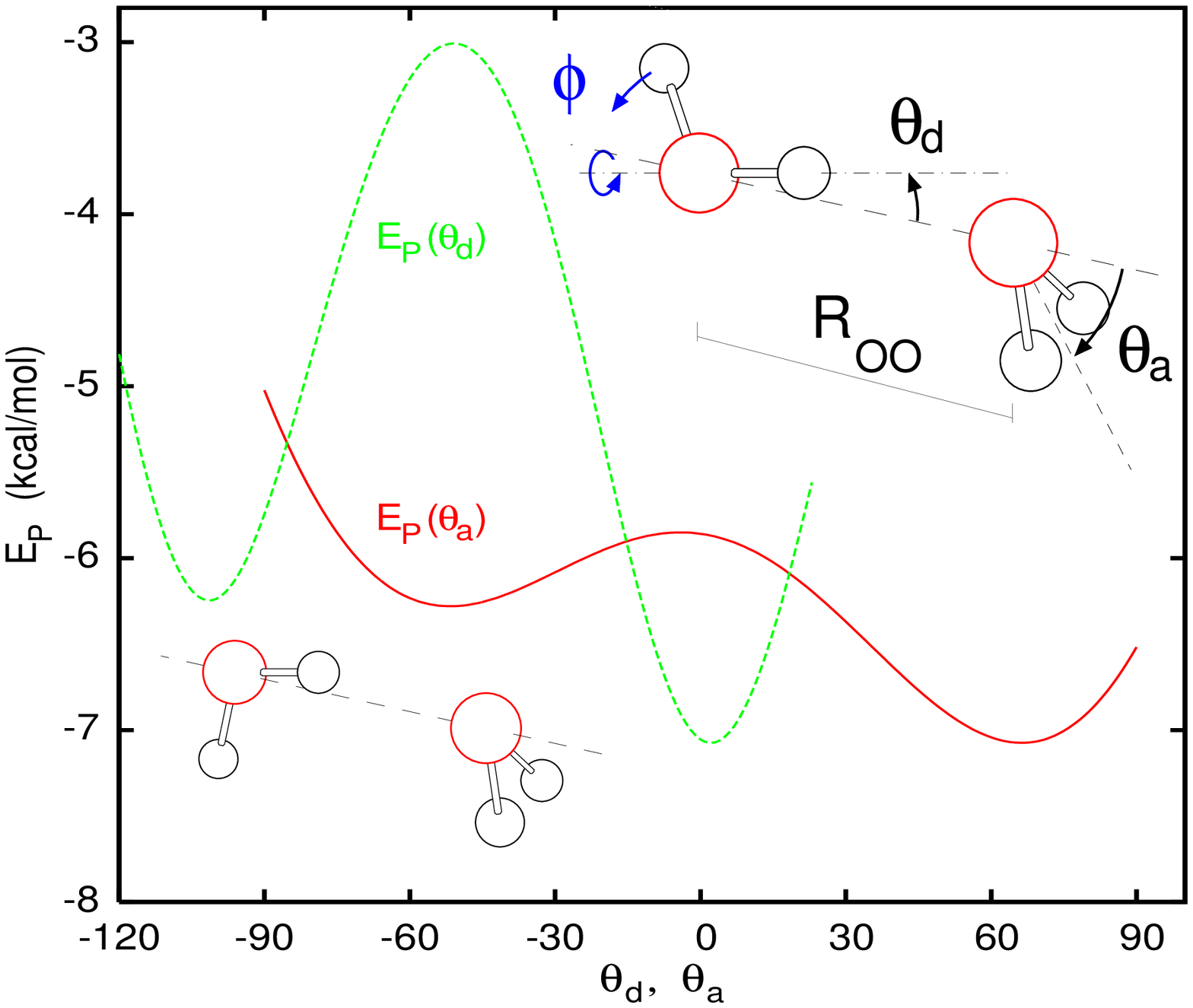,width=6.8cm,angle=-90}}
 \vskip 0.01 in
 \vbox{\baselineskip0.4 pt
 {\small {\bf Fig.~2} Polarization energy (Coulomb + induction),
$E_P(\theta_a,\theta_d)$ for water dimer.  $R_{OO}=2.977$\AA, and one
of the angles is fixed at the minimum value, as the other one is
varied.  $E_P$ is calculated for the defined system of atomic
multipoles and polarizabilities in the standard manner,\cite{pplus} by
computing fields of all multipoles of one molecule exerted on the
multipoles of another molecule, and solving for self-consistency. }}
 \vskip 0.1 in

Equilibrium hydrogen bonded configuration has a symmetry plane
(Fig.~2, inset) and is characterized by the oxygen-oxygen distance
$R_{OO}$, the donor angle $\theta_d$ and the acceptor angle
$\theta_a$.  The hydrogen bond forms when the donor proton points
against one of the lone pairs of the acceptor, $\theta_d\approx0$,
$\theta_a\approx\beta^{'}/2=54.74^\circ$ in our notation.  Actual
angles deviate slightly from these ideal values and are known with
some scatter.

For the experimental geometry we get $E_P=-7.046$ kcal/mol.  Adding
1.820 kcal/mol for the exchange and dispersion energy
[VRT(ASP-W)III\cite{VRTIII} value] we get equilibrium binding energy
$D_e=5.110$ kcal/mol.  The model also yields the total dipole moment
of the dimer in excellent agreement with experiment (Table I).

Since $E_P$ is only a part of the total interaction, which also
contains exchange and dispersion terms, we fix $R_{OO}$ and analyze
the orientation dependence (Fig.~2).  The minimum is achieved at
$\theta_d=2.14^\circ$ and $\theta_a=66.26^\circ$, which is close to,
but should not be confused with the equilibrium hydrogen-bonded
configuration, since other terms may shift the minimum.  Rotation of
either the donor by $\Delta\theta_d\approx-\beta$, or the acceptor by
$\Delta\theta_a\approx-\beta^{'}$ produces an alternative hydrogen
bonded arrangement sketched under the local minima in Fig.~2.

 \vskip 0.1 in
 \vbox{\baselineskip0.4 pt
 {\small
 {\bf Table I}.  Equilibrium binding energy $D_e$ (kcal/mol) and
dipole moment $\mu^{\text{dim}}$ (Debye) for water dimer.
$\mu_\bot^{\text{dim}}$ is the component of $\mu^{\text{dim}}$ normal
to the principal axis.  $^\dagger$Geometry is fixed at experimental
values; $^\ddagger$projection on the principal axis.\cite{VRT} }}

\centerline{
\begin{tabular}{l|ccc|lll}
\tableline
\tableline
      & $R_{OO}$ (\AA) & $\theta_d$ & $\theta_a$ & $D_e$      & $\mu^{\text{dim}}$ & $\mu_\bot^{\text{dim}}$ \\
\tableline
SAPT-5s & 2.955 & $6.36^\circ$ & $52.83^\circ$ & $4.858$ \\
SAPT-5st & 2.924 & $6.95^\circ$ & $58.52^\circ$ & $5.026$ \\
VRT(APS-W)III & 2.947 & $1.86^\circ$ & $49.27^\circ$ & $4.948$ & 2.69$^\ddagger$ \\
this work$^\dagger$ & 2.977 & $0.74^\circ$ & $59.7^\circ$  & $5.110$ & 2.67 & 0.13 \\
Expt.\cite{ExpGeomD,ExpQ} & 2.977 & $0.74^\circ$ & $59.7^\circ$  &         & 2.67 & 0.38 \\
\tableline
\tableline
\end{tabular}}
 \vskip 0.05 in

In order to further assess the quality of the model, we analyze the
energy variation along a path where the exchange and dispersion terms
vary little.  We choose to rotate the donor by an angle $\phi$ around
the bridging OH bond (Fig.~2).  Only a single proton then changes its
position and stays far from all the nuclei of the acceptor at all
$\phi$.

Figure 3 shows excellent agreement with all three best pair
potentials.  Note the small ($<1$ kcal/mol) total amplitude of the
variation, which is not described by a simple $\cos\phi$ function.
The overall agreement in the full range of $\phi$ is better with the
{\em ab-initio}-based SAPT-5s\cite{SAPT-5s} potential (the inset).
However, at small $\phi$ we get a near coincidence with the other two
curves, VRT(ASP-W)III and SAPT-5st,\cite{SAPT-5st} which are both
spectroscopically-tuned.  This is not surprising, assuming the
spectroscopic tuning is more sensitive to the region near the
equilibrium.

Explicit distributed polarizabilities (\ref{alpha2}) suggest an
estimate of the dispersion energy.  Due to the fast $r^{-6}$ decay,
the dispersion is dominated by two terms,
$E^D_{OO}\propto\alpha_O\alpha_O$ and
$E^D_{HO}\propto\alpha_H\alpha_O$.  Small $\alpha_H$ in the second
term is compensated by the proximity of the donor hydrogen to the
oxygen of the acceptor.  Neglecting dispersion nonadditivity and
assuming an universal scaling of the dispersion coefficient
$C_6\approx z\alpha_A\alpha_B$ for $A$ and $B$ species, we get
$E^D_{OO}=z\alpha_O^2/R_{OO}^6=0.99$ kcal/mol and
$E^D_{HO}=\frac{2}{3}z\alpha_H\alpha_O/(R_{OO}-l)^6=0.40$ kcal/mol for
linear hydrogen bond.  The total $E^D=1.39$ kcal/mol can be compared
to 1.56 kcal/mol from Fig.~3 of Ref.~\onlinecite{scoles_2001}.  For
this crude estimate we used $z=344$ kcal/mol value for Ar.  The factor
$\frac{2}{3}$ accounts for the anisotropy of $\alpha_H$.

\centerline{\epsfig{file=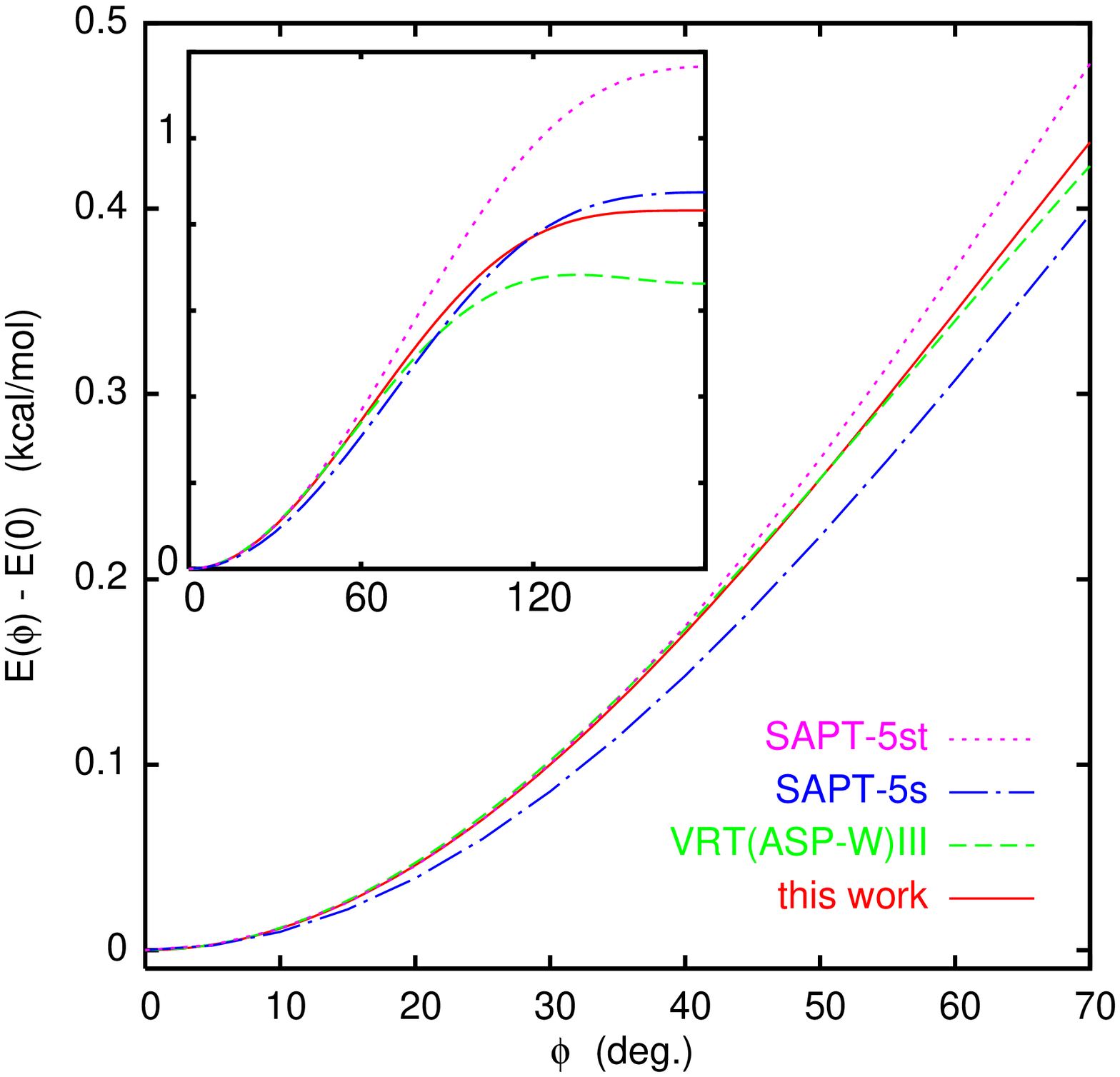,width=7cm,angle=-90}}

 \vbox{\baselineskip0.4 pt
 {\small {\bf Fig.~3} Energy variation for water dimer with rotation
of the donor around the bridging OH bond.  }}
 \vskip 0.05 in

Since the minimal model is constructed based solely on monomer
properties, we may speculate that it should describe larger clusters
as well, where the non-pairwise additivity of energy is
important.\cite{SAPT-5s,saykally_trimer} Such nonadditivity results
from self-consistency of all the induced moments in the
cluster,\cite{pplus} and may be relevant for the cooperativity of
hydrogen bonding in protein secondary structures.\cite{scheffler}

This work makes a step towards a chemical model for polarization
intermolecular forces by combining minimal atomic multipoles with
distributed polarizabilities, which together yield a transparent model
for polarization forces in water.  Its success raises a question of
broader applicability, especially to polarization and hydrogen bonding
in peptides and proteins, and in water-protein interactions.

The author is grateful to M.~Pederson and G.~Scoles for enlightening
conversations.  He also appreciates the VRT-III computer program
provided by N.~Goldman.  Numerous discussions with A.~Shabaev,
Al.~L.~Efros, and J.~Feldman are kindly acknowledged.  This work was
supported by the Office of Naval Research.

\end{multicols}


\begin{references}


\bibitem{iceDipole}
 C.A. Coulson and D. Eisenberg,
 Proc. Roy. Soc. Lond. A. {\bf 291}, 445 (1966).

\bibitem{parrinello}
 B. Chen, I. Ivanov, M.L. Klein, and M. Parrinello,
 Phys. Rev. Lett. {\bf 91}, 215503 (2003).

\bibitem{gen}
 Genesis 1:1-2

\bibitem{Hbook}
 S. Scheiner
 {\em Hydrogen Bonding.  A theoretical perspective},
 (Oxford University Press, Oxford, 1997).

\bibitem{mame}
 E.V. Tsiper and K. Burke,
 J. Chem. Phys. 120, 1153 (2004).

\bibitem{kutepov}
 {\em Water: structure, state, solvation.  Recent Achievements}
 (in Russian) ed. by A.M. Kutepov,
 (Science, Moscow, 2003).

\bibitem{VRT}
 R.S. Fellers, C. Leforestier, L.B. Braly, M.G. Brown, and R.J. Saykally,
 Science, {\bf 284}, 945 (1999).

\bibitem{VRTIII}
 N. Goldman, R.S. Fellers, M.G. Brown, L.B. Braly, C.J. Keoshian,
C. Leforestier, and R.J. Saykally,
 J. Chem. Phys. {\bf 116}, 10148 (2002).

\bibitem{SAPT-5s}
 G.C. Groenenboom, E.M. Mas, R. Bukowski, K. Szalewicz, P.E.S. Wormer,
and A. van der Avoird,
 Phys. Rev. Lett. {\bf 84}, 4072 (2000).

\bibitem{SAPT-5st}
 G.C. Groenenboom, P.E.S. Wormer, A. van der Avoird, E.M. Mas,
R. Bukowski, and K. Szalewicz,
 J. Chem. Phys. {\bf 113}, 6702 (2000).

\bibitem{batista}
 E.R. Batista, S.S. Xanthreas, and H. Jonsson,
 J. Chem. Phys. {\bf 112}, 3285 (2000).

\bibitem{stillinger}
 F.H. Stillinger and A. Rahman,
 J. Chem. Phys. {\bf 60}, 1545 (1974).

\bibitem{finney}
 J.L. Finney,
 J. of Mol. Liquids {\bf 90}, 303 (2001).

\bibitem{SPC}
 H.J.C. Berendsen, J.P.M. Postma, W.F. van Gunsteren, and J. Hermans,
 in {\em Intermolecular Forces}, ed. by B. Pullman (Reidel, Dordrecht,
1981), p. 331.

\bibitem{TIP5P}
 M.W. Mahoney and W.L. Jorgensen,
 J. Chem. Phys. {\bf 112}, 8910 (2000).

\bibitem{POL1}
 J. Caldwell, L.X. Dang, and P.A. Kollman,
 J. Am. Chem. Soc. {\bf 112}, 9144 (1990).

\bibitem{krogh}
 D.N. Bernardo, Y. Ding, K. Krogh-Jespersen, and R.M. Levy,
 J. Phys. Chem. {\bf 98}, 4180 (1994).

\bibitem{rick_A}
 S.W. Rick, S.J. Stuart, and B.J. Berne,
 J. Chem. Phys. {\bf 101}, 6141 (1994).


\bibitem{POL5}
 H.A. Stern, F. Rittner, B.J. Berne, and R.A. Friesner,
 J. Chem. Phys. {\bf 115}, 2237 (2001).

\bibitem{lamoureux_A}
 G. Lamoureux, A.D. MacKerell, and B. Roux,
 J. Chem. Phys. {\bf 119}, 5185 (2003).

\bibitem{ren_A}
 P.Y. Ren and J.W. Ponder,
 J. Phys. Chem. B {\bf 107}, 5933 (2003).

\bibitem{yu_A}
 H.B. Yu, T. Hansson, and W.F. van Gunsteren,
 J. Chem. Phys. {\bf 118}, 221 (2003).

\bibitem{ExpGeom}
 W.S. Benedict, N. Gailar, and E.K. Plyler,
 J. Chem. Phys. {\bf 24}, 1139 (1956).

\bibitem{pplus}
 E.V. Tsiper and Z.G. Soos,
 Phys. Rev. {\bf B64}, 195124 (2001).

\bibitem{ExpAlpha}
 W.F. Murphy,
 J. Chem. Phys. {\bf 67}, 5877 (1977).

\bibitem{Expmu}
 S.A. Clough, Y. Beers, G.P. Klein, and L.S. Rothman,
 J. Chem. Phys. {\bf 59}, 2254 (1973).

\bibitem{ExpQ}
 J. Verhoeven and A. Dymanus,
 J. Chem. Phys. {\bf 52}, 3222 (1970);
 The value $Q_{zz}$ depends on the choice of the origin of
coordinates, and is cited here relative to the oxygen.

\bibitem{ExpGeomD}
 J.A. Odutola and T.R. Dyke,
 J. Chem. Phys. {\bf 72}, 5062 (1980).

\bibitem{saykally_trimer}
 N. Pugliano and R.J. Saykally,
 Science {\bf 257}, 1937 (1992).


\bibitem{curtiss_1979}
 L.A. Curtiss, D.J. Frurip, and M. Blander,
 J. Chem. Phys. {\bf 71}, 2703 (1979).

\bibitem{MCY}
 O. Matsuoka, E. Clementi, and M. Yoshimine,
 J. Chem. Phys. {\bf 64}, 1351 (1976).

\bibitem{stone_waterD}
 C. Millot, J.-C. Soetens, M.T.C.M. Costa, M.P. Hodges, and
A.J. Stone,
 J. Phys. Chem. A {\bf 102}, 754 (1998).

\bibitem{scoles_2001}
 X. Wu, M.C. Vargas, S. Nayak, V. Lotrich, and G. Scoles,
 J. Chem. Phys. {\bf 115}, 8748 (2001).

\bibitem{pople}
 B.J. Smith, D.J. Swanton, J.A. Pople, H.F. Schaefer III, and
L. Radom,
 J. Chem. Phys. {\bf 92}, 1240 (1990).

\bibitem{scheffler}
 J. Ireta, J. Neugebauer, M. Scheffler, A. Rojo, and M. Galvan,
 J. Phys. Chem. A {\bf 107}, 1432 (2003).

\bibitem{comment}
 B3LYP hybrid density functional with aug-cc-pVTZ basis set yields
geometry $l=0.9619$\AA\ and $\beta=105.08^\circ$.  The calculation was
performed using GAUSSIAN 98 program by M.J. Frisch et al., (Gaussian
Inc, Pittsburgh, 1995).




\end{references}
\end{document}